\documentclass[10pt,conference]{IEEEtran}
\usepackage{amssymb,amsmath,amsfonts,bm}
\usepackage{graphicx}
\usepackage{subfigure}

\begin{document}

\title{Searching for low weight pseudo-codewords
}

\author{
\authorblockN{Michael Chertkov}
\authorblockA{Theoretical Division and Center for Nonlinear
Studies\\ LANL, MS B213, T-13, Los Alamos, NM 87545\\ {\tt\small
chertkov@lanl.gov}} \and
\authorblockN{Mikhail Stepanov}
\authorblockA{Department of Mathematics,
The University of Arizona\\
617 N. Santa Rita Ave., Tucson, AZ 85721\\
{\tt\small stepanov@math.arizona.edu}} }


\maketitle

\begin{abstract}
Belief Propagation (BP) and Linear Programming (LP) decodings of Low
Density Parity Check (LDPC) codes are discussed. We summarize
results of instanton/pseudo-codeword approach developed for analysis
of the error-floor domain of the codes. Instantons are special, code
and decoding specific, configurations of the channel noise
contributing most to the Frame-Error-Rate (FER). Instantons are
decoded into pseudo-codewords. Instanton/pseudo-codeword with the
lowest weight describes the largest Signal-to-Noise-Ratio (SNR)
asymptotic of FER, while the whole spectra of the low weight
instantons is descriptive of the FER vs SNR profile in the extended
error-floor domain. First, we describe a general optimization method
that allows to find the instantons for any coding/decoding. Second,
we introduce LP-specific pseudo-codeword search algorithm that
allows efficient calculations of the pseudo-codeword spectra.
Finally, we discuss results of combined BP/LP error-floor
exploration experiments for two model codes.
\end{abstract}

\section{Introduction}

Low-Density-Parity-Check (LDPC) codes \cite{63Gal,99Mac} are
special, not only because they can approach virtually error-free
transmission limit, but mainly because a computationally efficient
iterative decoding scheme,  the celebrated Belief Propagation (BP)
decoding, is readily available. For an idealized code on a tree, the
BP algorithm is exactly equivalent to the symbol-MAP decoding, which
is reduced to block-MAP (or simply Maximum Likelihood, ML), in the
asymptotic limit of infinite SNR. For any realistic code (with
loops), the BP algorithm is approximate, and it should actually be
considered as an algorithm solving iteratively nonlinear equations,
called BP equations. The BP equations describe extrema (e.g. minima
are of main interest) of the Bethe free energy \cite{05YFW}.
Minimizing the Bethe free energy, that is a nonlinear functional of
the probabilities/beliefs, under the set of linear (compatibility
and normalization) constraints, is generally a difficult task.

Linear Programming decoding, introduced in \cite{05FWK},  is a close
relative of BP which can be viewed as a relaxed version of Maximum
Likelihood (ML) decoding. Relation of the LP decoding to the Bethe
free energy approach \cite{05YFW}, and thus to BP equations and
decoding, was noticed in \cite{05FWK}, and the point was elucidated
further in \cite{03KV,04VK,05VK,06CS,06CCc,07CSa}. In short, LP may
be considered as large SNR asymptotic limit of BP, where the later
is interpreted as an extremum of the Bethe free energy functional.

Both BP and LP are computationally efficient but suboptimal,  i.e.
incapable of matching performance of the Maximum-Likelihood (ML).
Performance of an error-correcting scheme can be measured in terms
of the Frame Error Rate (FER) dependence on the Signal-to-Noise
Ratio (SNR). FER decreases as SNR increases. Even though BP and LP
decodings are suboptimal with respect to ML at all SNRs,  the
difference in FER is only order one in the water-fall regime of
small SNRs. The situation becomes significantly worse in the
error-floor domain of moderate to large SNRs where FER for BP/LP is
parametrically,  i.e. orders of magnitude, larger than FER for ML.
Length of the error-correction code brings another dimension into
the problem. The longer the code the lower is the value of FER where
the water-fall-to-error-floor transition happens. On the other hand,
standard Monte-Carlo (MC) numerics is incapable to determine BER
below $10^{-9}$. Therefore, understanding and describing the
error-floor by an alternative, and hopefully more insightful, method
is in great demand \cite{03Ric}.

One such useful insight came through recent efforts
\cite{04CCSV,05SCCV,05SC,06CS,07CSa} to understanding error-floor in
terms of the most probable of the dangerous configurations of the
noise, so-called instantons, contributing most to FER. BP/LP decodes
the instantons into the so-called non-codeword pseudo-codewords
\cite{95WLK,96Wib,01FKKR,03Ric,03KV}. It was recognized that for
moderate and large SNRs splitting of the two (FER vs SNR) curves,
representing ML decoding and approximate BP/LP decoding, is due to
the pseudo-codewords, which are confused by the suboptimal algorithm
for actual codewords of the code. Describing BP/LP error-floor
translates into finding pseudo-codewords with low effective
distance.

We discuss the instanton/pseudo-codeword approach in this
presentation. The two main themes reviewed are instanton-amoeba
\cite{05SCCV,06SCa} and LP-based Pseudo-Codeword Search (PCS)
algorithms \cite{06CS} for finding low effective distance
instantons/pseudo-codewords.

Instanton-amoeba, introduced in \cite{04CCSV,05SCCV}, is an
efficient numerical scheme which finds instanton/pseudo-codeword by
means of a simplex (amoeba) optimization. The algorithm is
initialized with a random simplex and many sequential attempts are
required to built the instanton/pseudo-codeword frequency spectra of
the code. The scheme is ab-initio by construction, thus it requires
no additional assumptions. It is also generic, in that there are no
restrictions related to the type of decoding or channel. The
instanton-amoeba method is general but also computational resources
consuming.

LP-based Pseudo-Codewords Search (PCS) algorithm, suggested in
\cite{06CS}, is an efficient alternative to the instanton-amoeba.
Formally,  one step of the PCS constitute sequential repetition of
the LP algorithm, where the entry information,  log-likelihoods, are
updated according to a feedback from the previous iteration in the
sequence. Like in the instanton-amoeba case, each step of the
algorithm starts from a randomly selected configuration of the noise
and ends at a low weight pseudo-codeword. PCS takes advantage of
some special features of LP resulted in monotonicity of the
procedure.  We experimentally observed that effective distance of
the result always decreases, or stays the same, after a single PCS
circle.

The two methods, instanton-amoeba and LP-based PCS,  can also be
viewed as complementary ingredients of one package aiming at
exploring the error-floor domain.  Thus, results of the PCS
algorithm can be naturally used as a starting guess for the
instanton-amoeba and vice-versa.

The material in the manuscript is organized as follows. Relation
between LP and BP decodings is elucidated, via the unifying Bethe
free energy approach, in Section \ref{sec:Bethe}. We introduce the
instanton-amoeba method in Section \ref{sec:amoeba}. LP-based PCS
algorithm is described in Section \ref{sec:PCS}. Simulation results
demonstrating utility of the instanton-amoeba and the PCS algorithms
are described in Section \ref{sec:results}. Here we discuss
$[155,64,20]$ and $[672,336,16]$ codes also presenting some Monte
Carlo simulations. Discussion of open problems, given in Section
\ref{sec:concl}, concludes the presentation.

\section{Belief Propagation and Linear Programming Decodings}
\label{sec:Bethe}

We consider a generic linear code, described by its parity check
$N\times M$ sparse matrix, $\hat{H}$, representing $N$ bits and $M$
checks. The codewords are configurations,
${\bm\sigma}=\{\sigma_i=0,1| i=1,\dots,N\}$, which satisfy all the
check constraints: $\forall \alpha=1,\dots,M$, $\sum_i H_{\alpha
i}\sigma_i=0$ (mod~$2$). The codeword sent to the channel is
polluted and the task of decoding becomes to restore the most
probable pre-image of the output sequence,  ${\bm x}=\{x_i\}$.
Probability for ${\bm\sigma}$ to be a pre-image of ${\bm x}$ is
\begin{equation}
{\cal P}({\bm\sigma}|{\bm x})\!=\!Z^{-1}\prod_\alpha
\delta\biggl(\prod_{i\in\alpha}(-1)^{\sigma_i},1\biggr)
\exp\biggl(-\sum_ih_i\sigma_i\biggr), \label{Psx}
\end{equation}
where one writes $i\in\alpha$ if $H_{\alpha i}=1$; $Z$ is the
normalization coefficient (so-called partition function); the
Kronecker symbol, $\delta(x,y)$, is unity if $x=y$ and it is zero
otherwise; and ${\bm h}$ is the vector of log-likelihoods dependent
on the output vector ${\bm y}$. In the case of the AWGN channel with
the SNR ratio, $SNR=E_c/N_0=s^2$, bit transition probability is,
$\sim \exp(-2s^2(x_i-\sigma_i)^2)$, and the log-likelihood becomes,
$h_i=s^2(1-2x_i)$. The optimal block-MAP (Maximum Likelihood)
decoding maximizes ${\cal P}({\bm\sigma}|{\bm x})$ over
${\bm\sigma}$
\begin{equation}
\arg\max_{\bm\sigma}{\cal P}({\bm\sigma}|{\bm x}), \label{ML}
\end{equation}
and symbol-MAP operates similarly, however in terms of the marginal
probability at a bit
\begin{equation}
\arg\max_{\sigma_i}\sum_{{\bm\sigma}\setminus\sigma_i}{\cal
P}({\bm\sigma}|{\bm x}). \label{sMAP}
\end{equation}

BP and LP decodings should be considered as computationally
efficient but suboptimal substitutions for MAP decodings. Both
decodings can be conveniently derived from the so-called Bethe-Free
energy approach of \cite{05YFW} which is briefly reviewed below.
(See also \cite{06CCb,06CCc,07CC}.) In this approach trial
probability distributions, called beliefs, are introduced both for
bits and checks, $b_i$ and $b_\alpha$, respectively. The set of
bit-beliefs, $b_i(\sigma_i)$, satisfy equality and inequality
constraints that allow convenient reformulation in terms of a bigger
set of beliefs defined on checks, $b_\alpha({\bm\sigma}_\alpha)$,
where, ${\bm \sigma}_\alpha=\{\sigma_i|i\in\alpha,\sum_i H_{\alpha
i}\sigma_i=0\mbox{ (mod~$2$)}\}$, is a local codeword associated
with the check $\alpha$. The equality constraints are of two types,
normalization constraints (beliefs, as probabilities, should sum to
one) and compatibility constraints
\begin{equation}
\forall i,\ \forall\alpha\ni i:\
b_i(\sigma_i)=\sum\limits_{\sigma_\alpha\setminus\sigma_i}b_\alpha({\bm
\sigma}_\alpha),\
\sum\limits_{{\bm\sigma}_\alpha}b_\alpha({\bm\sigma}_\alpha)=1.\label{comp}
\end{equation}
Additionally all the beliefs should be non-negative and smaller than
or equal to unity. The Bethe Free energy is defined as a difference
of the self-energy and the entropy, $F=E-S$:
\begin{eqnarray}
 && \!\!\!\!\!\!\!\!\!\!\!\!\!\!\!\!\!\!\!\!\!\!\!
 E\!=\sum_ih_i\sum_{\sigma_i}\!\sigma_ib_i(\sigma_i),\label{E_Bethe}\\
 && \!\!\!\!\!\!\!\!\!\!\!\!\!\!\!\!\!\!\!\!\!\!\!
  S\!\!=\!-\!\!\sum\limits_\alpha\! \sum_{{\bm \sigma}_\alpha}\!
  b_\alpha\!({\bm\sigma}_\alpha\!)
 \!\ln b_\alpha\!({\bm\sigma}_\alpha\!)\!
 +\!\!\sum\limits_i\!\sum\limits_{\sigma_i}
 (q_i\!-\!1)b_i(\sigma_i)\!\ln\! b_i(\sigma_i).
 \label{S_Bethe}
\end{eqnarray}
Optimal configurations of beliefs minimize the Bethe Free energy
subject to the equality constraints (\ref{comp}). Introducing the
constraints as the Lagrange multiplier terms to the effective
Lagrangian and looking for the extremum with respect to all possible
beliefs leads to
\begin{eqnarray}
 &&\!\!\!\!\!\!\!\!\!\! b_\alpha({\bm
 \sigma}_\alpha)=\frac{\exp\left(\sum_{i\in\alpha}(h_i/q_i+\eta_{\alpha
 i})(1-2\sigma_i)\right)}{\sum_{{\bm\sigma}_\alpha}\exp\left(\sum_{i\in\alpha}(h_i/q_i+\eta_{\alpha
 i})(1-2\sigma_i)\right)},\label{ba}\\
 &&\!\!\!\!\!\!\!\!\!\! b_i(\sigma_i)=\frac{\exp\left((\eta_{\alpha
 i}+\eta_{i\alpha})(1-2\sigma_i)\right)}{2\cosh\left(\eta_{i\alpha}+\eta_{\alpha
 i}\right)},\label{bi}
\end{eqnarray}
where the set of $\eta$ fields (which are Lagrange multipliers for
the compatibility constraints) satisfy
\begin{eqnarray}
 \eta_{\alpha i}=h_i+\sum_{\beta\ni i}^{\beta\neq \alpha}\eta_{i\alpha},\ \
 \eta_{i\alpha}=\tanh^{-1}\left(\prod_{j\in\alpha}^{j\neq
 i}\tanh\eta_{\alpha j}\right).
 \label{eta}
\end{eqnarray}
These are the BP equations for LDPC codes written in its standard
form. These equations are often described in the coding theory
literature as stationary point equations for the BP (also called sum
product) algorithm and then $\eta$ variables are called messages.
The BP algorithm, initialized with $\eta_{i\alpha}=0$ and iterating
Eqs.~(\ref{eta}) sequentially from right to left, is exact on the
tree. This is the BP algorithm discussed in the paper. However,  let
us notice  for the sake of completeness,  that even though
significance of the BP equations for MAP decoding of actual codes
(with loops) was established \cite{06CCa,06CCb,06CCc}, the standard
tree-motivated choice of the BP iterations scheduling is not
obvious.  Possible lack of the iterative algorithm convergence (to
respective solution of the BP equation) is a particular concern, and
some relaxation methods were recently introduced to deal with the
problem \cite{06SCb,07LHMH}.

LP is a close relative of BP which does not have this unpleasant
problem with convergence. Originally, LP decoding was introduced as
a relaxation of ML decoding \cite{05FWK}. Eq.~(\ref{ML}) can be
restated as
\begin{equation}
\arg\min_{{\bm\sigma}\in {\cal P}}\biggl(\sum_ih_i\sigma_i\biggr),
\label{ML-LP}
\end{equation}
where ${\cal P}$ is the polytope spanned by all the codewords of the
code. Looking for ${\bm\sigma}$ in terms of a linear combination of
the codewords, ${\bm \sigma}_v$: ${\bm\sigma}=\sum_v\lambda_v{\bm
\sigma}_v$, where $\lambda_v\geq 0$ and $\sum_v\lambda_v=1$, one
observes that block-MAP turns into a linear optimization problem.
LP-decoding algorithm of \cite{05FWK} proposes to relax the
polytope, expressing ${\bm\sigma}$ in terms of a linear combination
of local codewords associated with checks, ${\bm\sigma}_\alpha$. We
will not give details of this original formulation of LP here,
because we prefer an equivalent formulation elucidating connection
to BP decoding. One finds that BP decoding, understood as an
algorithm searching for a stationary point of the BP equations,
turns into LP decoding in the asymptotic limit of large SNR. Indeed
in this special limit the entropy terms in the Bethe free energy can
be neglected and the problem turns to minimization of a linear
functional under a set of linear constraints. The similarity between
LP and BP (the later one identified with a minimum of the Bethe Free
energy \cite{05YFW}) was noticed in \cite{05FWK} and it was also
discussed in \cite{03KV,04VK,05VK,06CCc}. Stated in terms of
beliefs, LP decoding minimizes the self-energy part (\ref{E_Bethe})
of the full Bethe Free energy functional under the set of linear
equality constraints (\ref{comp}) and also linear inequalities
guaranteeing that all the beliefs are non-negative and smaller than
or equal to unity. This gives us full definition of the so-called
large polytope LP decoding. One can run it as is in terms of bit-
and check- beliefs, however it may also be useful to re-formulate
the LP procedure solely in terms of the bit beliefs. The small
polytope formulation of LP is due to \cite{91Yan} and \cite{05FWK}.
Indeed, self-energy is stated only in terms of bit beliefs, and
moreover one rewrites it just in terms of $f_i=b_i(1)$, excluding
$b_i(0)=1-f_i$ from the consideration. Furthermore, one can also
exclude check beliefs, replacing them by a set of inequality
constraints imposed on $f_i$. The later remain the only set of
variables stayed in the small polytope formulation,
$\forall\alpha,\forall T\subseteq {\cal
N}(\alpha)=\{i;i\in\alpha\},\
 |T|\mbox{ is odd }$:
\begin{equation}
 \sum_{i\in T} f_i+\sum_{i\in(N(\alpha)\setminus T)}(1-f_i)\leq
 |N(\alpha)|-1.
\label{small}
\end{equation}

\section{Instanton-amoeba as a general method of the noise space
exploration \cite{05SCCV,06SCa}} \label{sec:amoeba}

\begin{figure}[b]
  \centerline{\includegraphics[width=3in]{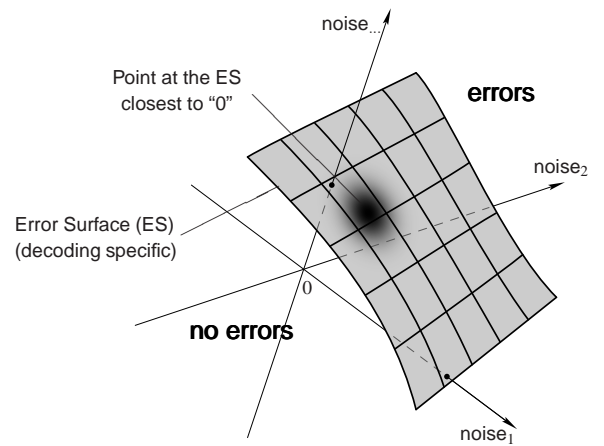}}
  \caption{Illustration for the instanton method. The noise space is
    divided into areas of successful and erroneous decoding by error
    surface. The point at the error surface closest (in the appropriate
    metrics) to the point of zero noise is
    the most probable configuration of the noise causing the decoding error.
    Contribution from the special configuration of the noise, the instanton,
    and its close vicinity estimates the noise integral for FER.}
  \label{inst}
\end{figure}

Goal of decoding is to infer the original message from the received
output, ${\bm x}$. Assuming that coding and decoding are fixed and
aiming to characterize performance of the scheme, one studies
Frame-Error-Rate (FER)
 $\mbox{ FER} = \int d{\bm x} \,\, \chi_{\mbox{\scriptsize error}}({\bm
   x}) P({\bm x}|{\bm 0})$,
where $\chi_{\mbox{\scriptsize error}} = 1$ if an error is detected
and $\chi_{\mbox{\scriptsize error}} = 0$ otherwise. In symmetric
channel FER is invariant with respect to the original codeword, thus
all-$0$ codeword can be assumed for the input. When SNR is large
FER, as an integral over output configurations, is approximated by,
\begin{eqnarray}
\mbox{FER}\sim \sum\limits_{\mbox{\scriptsize inst}}
V_{\mbox{\scriptsize inst}} \times P\left({\bm x}_{\mbox{\scriptsize
inst}}|{\bm 0}\right), \label{FER}
\end{eqnarray}
where ${\bm x}_{\mbox{\scriptsize inst}}$ are the special instanton
configurations of the output maximizing $P({\bm x}|{\bm 0})$ under
the $\chi_{\mbox{\scriptsize error}} = 1$ condition, and
$V_{\mbox{\scriptsize inst}}$ combines combinatorial and
phase-volume factors. See Fig.~\ref{inst} for illustration.
Generally, there are many instantons that are all local maxima of
$P({\bm x}|{\bm 0})$ in the noise space.

For the AWGN channel (considered as the main model example) finding
the instanton means minimizing $d = {\bm x}^2$ with respect to the
noise vector ${\bm x}$ in the $N$-dimensional space and under the
condition that the decoding terminates with an error. Instanton
estimation for FER at the highest SNR, $s \gg 1$, is $\sim
\exp(-d_{\mbox{\scriptsize min}} \cdot s^2/2)$, while at moderate
values of SNR many terms from the right-hand-side of Eq.~(\ref{FER})
can contribute to FER comparably.

In our instanton-amoeba numerical scheme instanton with the smallest
effective distance, $d_{\mbox{\scriptsize min}}$, was found by a
downhill simplex method also called ``amoeba'', with accurately
tailored (for better convergence) annealing. We repeat the
instanton-amoeba evaluation many times, always starting from a new
set for initial simplex chosen randomly. $d$,  as a function of
noise configuration inside the area of unsuccessful decoding, has
multiple minima each corresponding to an instanton. Multiple
attempts of the instanton-amoeba evaluations gives us not only the
instanton with the minimal $d_{\mbox{\scriptsize inst}}$ but also
the whole spectra of higher valued $d_{\mbox{\scriptsize inst}}$.

Instanton is a highly probable configuration of the noise leading to
an error. Decoding applied to the instanton configuration results in
the so-called pseudo-codeword \cite{95WLK,96Wib,01FKKR,03Ric,03KV}.
Effective distance, $d_{\mbox{\scriptsize inst}}$, characterizing an
instanton and its respective pseudo-codeword, should be compared
with the Hamming distance of the code, $d_{\mbox{\scriptsize ML}}$,
which measures minimal number of flips (from $0$ to $1$ and vice
versa) required for changing from the all zero codeword to another
codeword of the code. Instanton/pseudo-codeword with
$d<d_{\mbox{\scriptsize ML}}$ will completely screen contribution of
the respective codeword into FER at the largest SNRs. We will see
below in Section \ref{sec:results}) that this situation is actually
realized for one of the codes discussed.

We develop two different versions of ``amoeba",  ``soft" and
``hard". In ``soft amoeba" the minimization function decreases with
noise probability density in erroneous area of the noise, while in
area of successful decoding the function is made artificially big
(to guarantee that the actual minimum is achieved inside the
erroneous domain). In the ``hard amoeba'' case minimization is
performed only over all orientations of the noise vector, while the
length of the vector corresponds exactly to respective point at the
error surface, that is the surface separating domains of errors from
the domain of correct decoding. (See Fig.~\ref{inst} for
illustration.) This special point at the error surface is found
numerically by bisection method. In \cite{05SCCV,05SC} the ``hard
amoeba" was used. In \cite{06SCa} we found that even though the
``hard amoeba" outperforms the ``soft amoeba" for relatively short
codes,  the later one has clear advantage in the computational
efficiency for mid-size and long codes.

Once an instanton is found numerically, its validity can be verified
against a theoretical evaluation. This theoretical approach,
introduced in \cite{05SCCV,05SC}, is based on the notion of the
computational tree (CT) of Wiberg \cite{96Wib} built by unwrapping
the Tanner graph of a given code into a tree from a bit for which
one determines the probability of error. The concept of CT is useful
because the result of iterative decodings at a bit of an LDPC code
and at the tree center of the respective CT are equal by
construction \cite{96Wib}. The initial messages at any bit of the
tree are log-likelihoods and, therefore, the result obtained in the
tree center is a linear combination of the log-likelihoods with
integer coefficients, so the error surface condition becomes $\sum_i
n_i h_i = 0$ with integer $n_i$ that depend on CT structure. For
AWGN channel the instanton length is equal to $d_{\mbox{\scriptsize
inst}} = (\sum_i n_i)^2/(\sum_i n_i^2)$ \cite{96Wib}. The definition
of $n_i$ was generalized in \cite{05SCCV}. In spite of its clear
utility the CT approach becomes impractical for larger number of
iterations. Thus, we actually use the CT approach only to verify
validity of the instanton-amoeba results for relatively small number
of iterations.


\section{Accelerated Pseudo-codeword search for LP decoding \cite{06CS}}
\label{sec:PCS}

\begin{figure}[t]
\centerline{\includegraphics[width=3.1in]{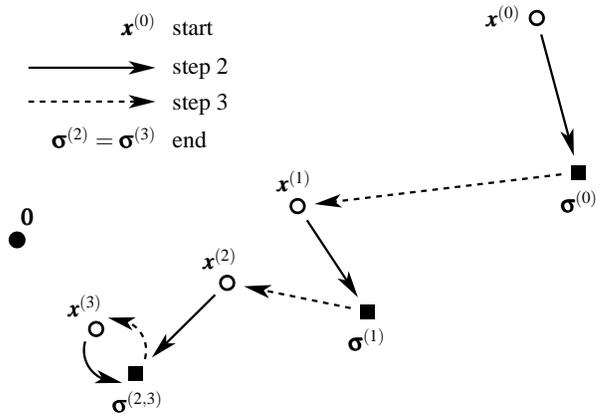}}
\caption{Schematic illustration of the pseudo-codeword-search
algorithm. This example terminates at $k_*=3$.} \label{fig:scheme}
\end{figure}

Suppose a pseudo codeword, $\tilde{\bm
\sigma}=\{\tilde{\sigma_i}=b_i(1);\ i=1,\dots,N\}$, corresponding to
the most damaging configuration of the noise (instanton) counted
from the all zero codeword, ${\bm x}_{\mbox{\scriptsize inst}}$, is
found. Then finding the instanton configuration itself (i.e.
respective configuration of the noise) is not a problem, one only
needs to maximize the transition probability with respect to the
noise field, ${\bm x}$, taken at ${\bm \sigma}=0$ under the
condition that the self-energy calculated for the pseudo-codeword in
the given noise field ${\bm x}$ is zero (i.e. equal to the value of
the self energy for the zero code word). The resulting expression
for the optimal configuration of the noise (instanton) in the case
of the AWGN channel is
 ${\bm x}_{\mbox{\scriptsize inst}}=(\tilde{\bm
 \sigma}\sum_i\tilde{\sigma}_i)/(2\sum_i\tilde{\sigma}_i^2)$,
and the respective effective distance is
 $d_{\mbox{\scriptsize LP}}=(\sum_i\tilde{\sigma}_i)^2/
 \sum_i\tilde{\sigma}_i^2$.
This definition of the effective distance was first described in
\cite{01FKKR},  with the first applications of this formula to LP
decoding discussed in \cite{03KV} and \cite{05VK}. Note also that
the expressions are reminiscent of the formulas derived by Wiberg
and co-authors in \cite{95WLK} and \cite{96Wib}, in the context of
the computational tree analysis applied to iterative decoding with a
finite number of iterations.

Let us now introduce the pseudo-codeword-search algorithm
\cite{06CS} inspired by the aforementioned median procedure.
\begin{itemize}
\item
{\bf Start:} Initiate a starting configuration of the noise, $ {\bm
x}^{(0)}$. Noise is counted from zero codeword and it should be
sufficiently large to guarantee convergence of LP to a
pseudo-codeword different from the zero codeword.
\item
{\bf Step 1:} LP decodes $ {\bm x}^{(k)}$ to a codeword ${\bm
\sigma}^{(k)}$.
\item
{\bf Step 2:} Find ${\bm y}^{(k)}$, the weighted median in the noise
space between the pseudo codeword, ${\bm \sigma}^{(k)}$, and the
zero codeword.  The AWGN expression for the weighted median is
 ${\bm y}^{(k)}=({\bm \sigma}^{(k)}\sum_i\sigma_i^{(k)})/(2
 \sum_i\big(\sigma_i^{(k)}\big)^2)$.
\item
 {\bf Step 3:}
 If ${\bm y}^{(k)}={\bm y}^{(k-1)}$, then $k_*=k$ and the algorithm
terminates. Otherwise go to Step 2, assigning ${\bm x}^{(k+1)}={\bm
y}^{(k)}+0$. ($+0$ prevents decoding into the zero codeword, keeping
the result of decoding within the erroneous domain.)
\item
{\bf Output} configuration ${\bm y}^{(k_*)}$ is the configuration of
the noise that belongs to the error-surface surrounding the zero
codeword. (The error-surface separates the domain of correct LP
decisions from the domain of incorrect LP decisions.) Moreover,
locally,  i.e. for the given part of the error-surface equidistant
from the zero codeword and the pseudo codeword ${\bm
\sigma}^{(k_*)}$, ${\bm y}^{(k_*)}$ is the nearest point of the
error-surface to the zero codeword.
\end{itemize}

We repeat the algorithm many times picking the initial noise
configuration randomly,  however guaranteeing that it would be
sufficiently far from the zero codeword so that the result of the LP
decoding (first step of the algorithm) is a pseudo-codeword distinct
from the zero codeword. We showed in \cite{06CS} that the PCS
algorithm converges in a relatively small number of iterations.

For the sake of completeness, let us also notice that there are some
LP-specific limitations which are carried over to the bare PCS
algorithm. Thus, LP decoding operates with the local codewords while
their number grows exponentially with check degrees, $q_\alpha$.
However, this undesirable complexity of LP can actually be dealt
with. It was noticed in \cite{05FWK} that only relatively few of the
LP constraints are actually used in decoding. Some suggestions were
introduced to overcome the problem \cite{05FWK,06TS,06VK,07CSa}.
Even though any of the complexity reduction method can probably be
used to improve performance of PCS, our only tests so far were based
on the dendro-LDPC method of \cite{07CSa}. The dendro-LDPC approach
suggests to change the graphical representation of the model by
replacing all checks of high degree by dendro-subgraphs (trees) with
appropriate number of auxiliary checks of degree three and number of
punctured, i.e. not transmitted, bits of degree two. We showed in
\cite{07CSa} that the dendro-code and the original code have
identical set of codewords and pseudo-codewords. Moreover,  for any
configuration of the channel output the results of MAP decodings are
identical for the two codes. Another result,  reported in
\cite{07CSa}, is that the described above PCS algorithm works
flawlessly for the dendro-codes. The dendro version of the algorithm
is actually identical to the one described above under exception of
what concerns the punctured nodes. First, one should always zero the
log-likelihoods at all the punctured nodes and, second, calculating
the weighted medians one should exclude punctured nodes from the
sum.

Our direct attempts to extend the PCS algorithm to BP decoding did
not succeed. In this regards, we attribute success of the PCS in the
LP case to the fact that the weighted median ($+0$) of the zero
codeword and a pseudo codeword {\it is not} decoded into the zero
codeword, generating a new pseudo-codeword with effective distance
smaller or equal to (but never larger than!) the one of the initial
pseudo-codeword. This monotonicity feature is apparently lucking in
the standard iterative BP. In spite of that we still found an
indirect way of using the PCS LP results for analysis of the BP
decoding. One simply uses result of the LP-PCS as entry guess for
the BP instanton-amoeba search. The hybrid method works well, often
resulting in discovery of BP instantons/pseudo-codewords with small
effective distance.

\section{Pseudo-codeword spectra and FER vs SNR performance curve}
\label{sec:results}

\begin{figure}[t]
  \centerline{\includegraphics[width=3in]{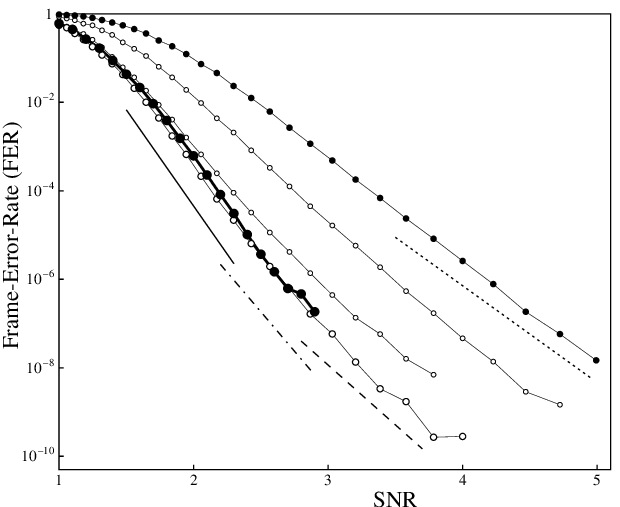}}
  \medskip
  \centerline{\includegraphics[width=2.9in]{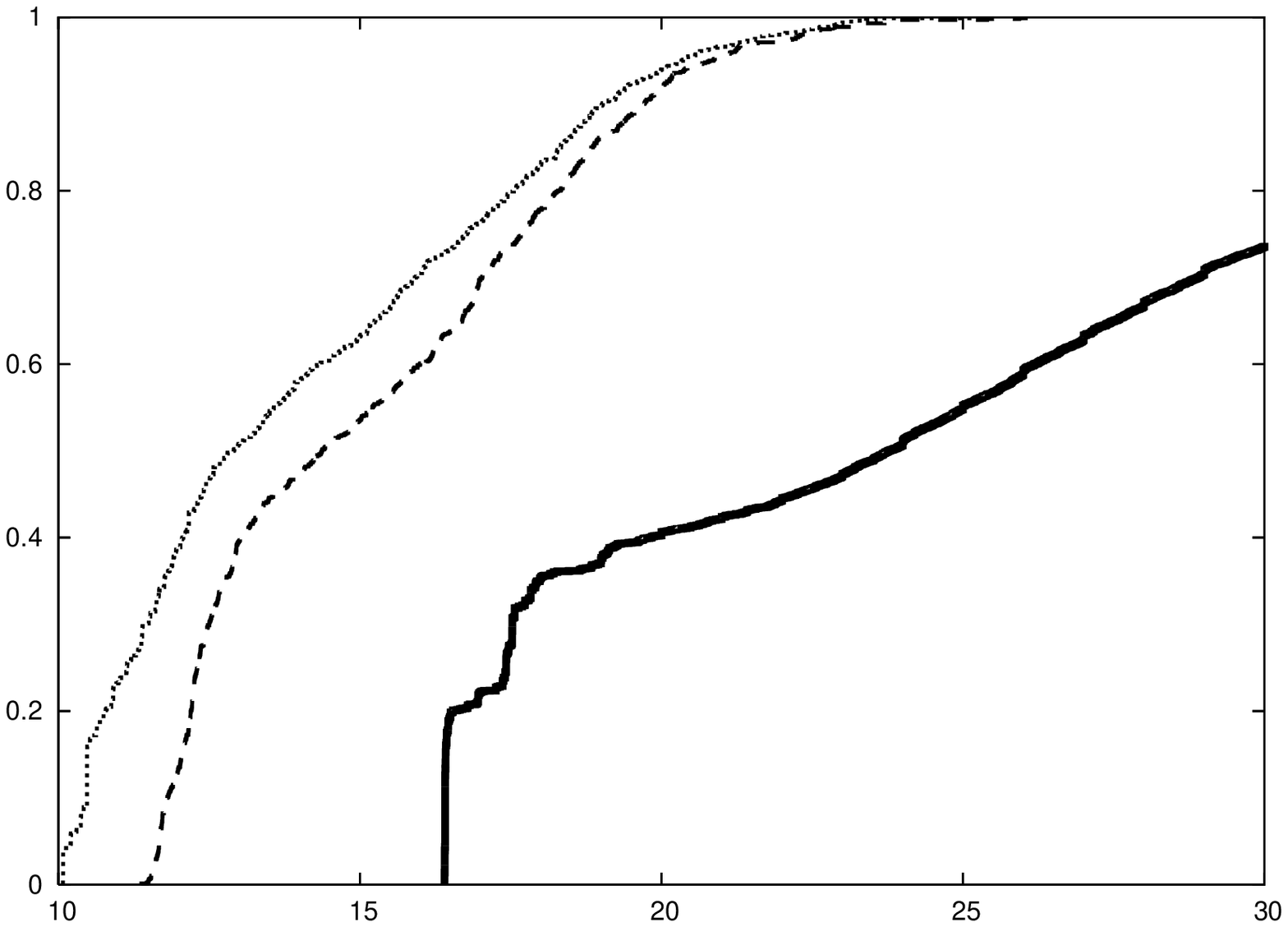}}
\caption{FER vs SNR ($=s^2$) curves for the $[155,64,20]$ code
decoded by iterative BP decoding with $4,8,128,1024$ iterations and
LP decoding are shown on the top Figure A. Filled dots for the data
points correspond to $4$ iterations BP decoding. Improvement of the
code performance with the number of iterations is monotonic. Filled
dots over bold line correspond to LP decoding. Solid straight line
corresponds to the Hamming distance asymptotics, $\mbox{FER} \sim
\exp(-20 \cdot s^2/2)$. Dotted straight line corresponds to the
minimal distance instanton for BP with $4$ iterations, $\mbox{FER}
\sim \exp(-d_4 \cdot s^2/2)$, where $d_4 = 46^2/210$. Dashed
straight line correspond to a special instanton configuration with
$d=12.5$ that withstand $~400$ BP iterations. Dash-dotted straight
line corresponds to the minimal distance instanton for LP decoding,
$\mbox{FER} \sim \exp(-d_{LP} \cdot s^2/2)$, where $d_{LP}\approx
16.4037$.
\newline  Figure B, on the
bottom, shows pseudo-codeword spectra found for iterative BP and LP
decodings by the instanton-amoeba and the PCS methods respectively.
Solid, dotted and dashed curve show results for the LP and BP
decodings with four and eight iterations respectively.
  }
  \label{fig:Tanner}
\end{figure}

\begin{figure}[t]

\centerline{\includegraphics[width=3in]{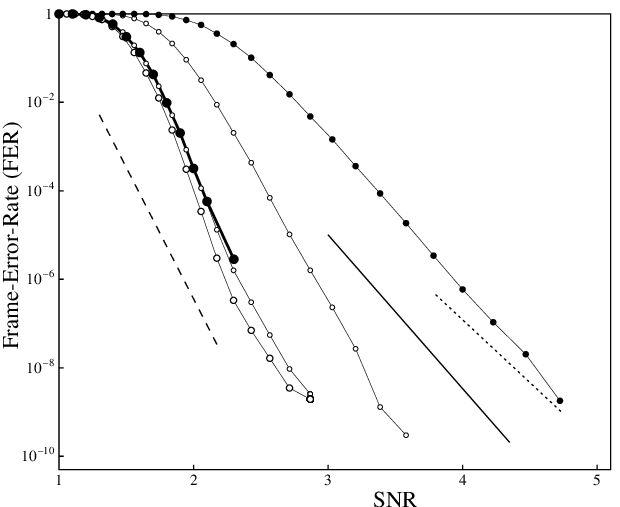}}
\medskip
\centerline{\includegraphics[width=2.9in]{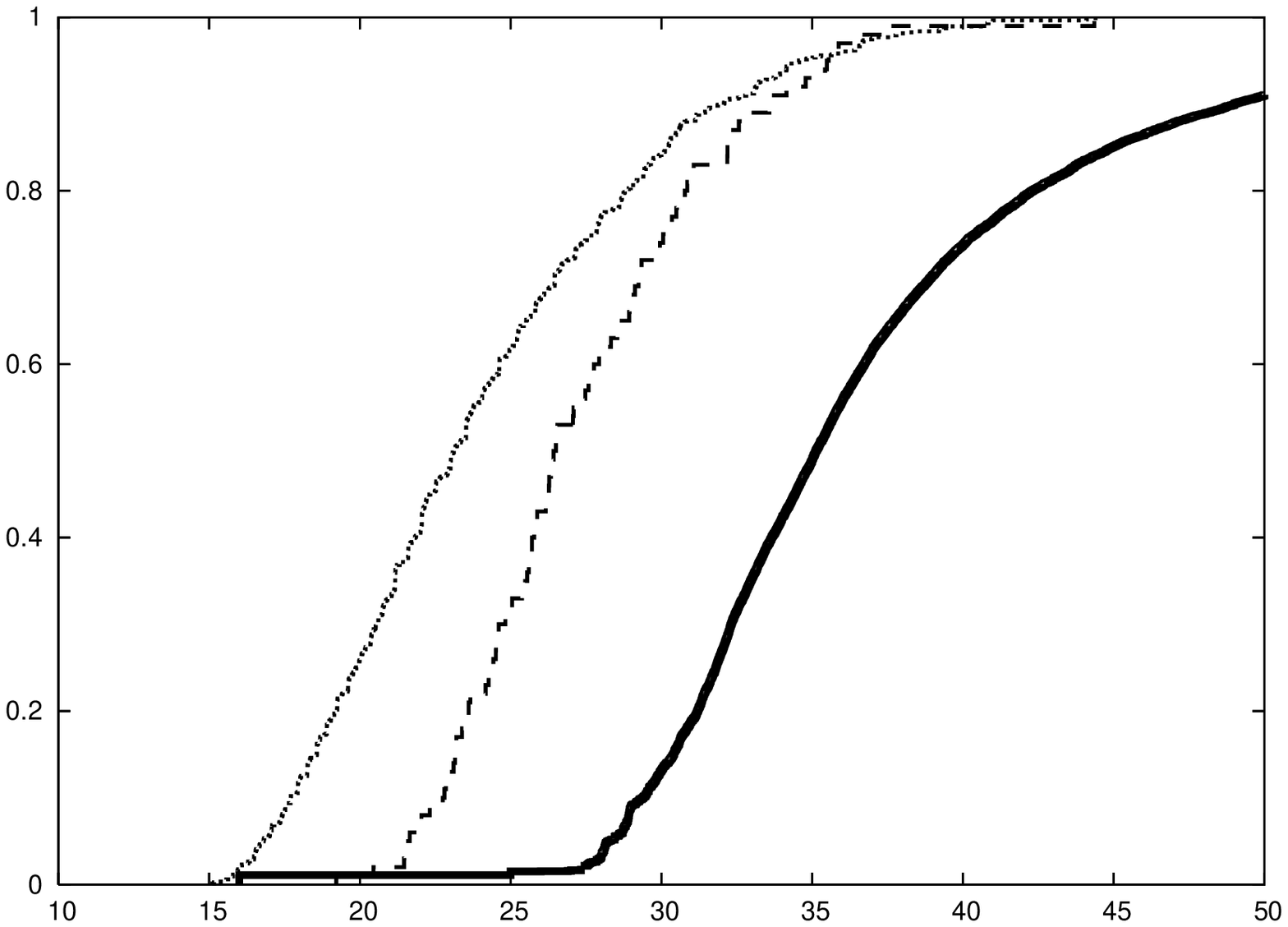}} \caption{FER
vs SNR ($=s^2$) curves for the $[672,336,16]$ code decoded by
iterative BP and LP decodings are shown on the top Figure A.
Different curves correspond to BP decoding with $4$, $8$, $128$ and
$1024$ iterations. Filled dots for the data points correspond to $4$
iterations decoding. Improvement of the code performance with the
number of iterations is monotonic. Filled dots over bold line
correspond to LP decoding. Solid straight line corresponds to the
Hamming distance asymptotics, $\mbox{FER} \sim \exp(-16 \cdot
s^2/2)$. Dotted straight line corresponds to the minimal distance
instanton for BP with $4$ iterations, $\mbox{FER} \sim
\exp(-d_{4;\mbox{\scriptsize min}} \cdot s^2/2)$, where
$d_{4;\mbox{\scriptsize min}} = 46^2/162$. Dashed straight line
corresponds to the minimal distance instanton for LP decoding,
$\mbox{FER} \sim \exp(-d_{\mbox{\scriptsize LP};\mbox{\scriptsize
min}} \cdot s^2/2)$, where $d_{LP;\mbox{\scriptsize min}}\approx
27.33$. \newline Figure B, on the bottom, shows pseudo-codeword
spectra,  i.e. probability to observe pseudo-codeword of a given or
smaller effective distance, found for iterative BP and LP decodings
by the instanton-amoeba and the PCS methods respectively. Solid,
dotted and dashed curve show results for the LP and  BP decodings
with four and eight iterations respectively.} \label{fig:p7}
\end{figure}

In this Section aiming to illustrate utility of the instanton-amoeba
and PCS methods described above, we analyze two codes, the Tanner
$[155,64,20]$ code introduced in \cite{01TSF}, and the $p=7$
Margulis code $[672,336,16]$ introduced in \cite{82Mar} and also
discussed in \cite{02MP}. The two codes are selected for
demonstration,  in part, because they show qualitatively different
behavior in the error-floor domain. The results discussed in this
Section were partially presented before in \cite{06SCa} and
\cite{06CS} for BP and LP decodings respectively.

We perform numerical simulations of three distinct and complementary
types.   First of all we study FER vs SNR curve for iterative BP
decoding (with fixed number of iterations) and LP decoding by direct
Monte-Carlo simulations. These MC results,  shown in
Fig.~\ref{fig:Tanner}A and Fig~\ref{fig:p7}A for the two codes
respectively, provide a test ground for the two other
instanton-amoeba and PCS methods aimed at exploring efficiently the
error-floor domain. The outcome of the error-floor exploration
experiment is contained in a list of observed low-weight
pseudo-codeword configurations,  that is compactly presented in the
form of the effective-distance spectra of the codes,  shown in
Fig.~\ref{fig:Tanner}B and Fig~\ref{fig:p7}B respectively.

In the case of the $[155,64,20]$ code the pseudo-codeword spectrum
of LP starts at $d_{\mbox{\scriptsize min}}\approx 16.404$ and the
pseudo-codeword frequency with the effective distance increase, e.g.
passing though $d_{\mbox{\scriptsize ML}}=20$ without any visible
anomaly. The growth starting immediately at $d_{min}$ is fast,
indicating that the frequency of the low-effective distance
configurations is considerable, i.e. $O(1)$. This form of the
pseudo-codeword spectra is fully consistent with what is seen in the
MC simulations: the error-floor asymptotic of FER, $\sim
\exp(-d_{\mbox{\scriptsize min}}s^2/2)$, correspondent to the
pseudo-codeword with the lowest effective weight, sets early. The
pseudo-codeword spectra of BP are qualitatively similar to the one
of LP. We anticipate that the BP spectra gets closer to the LP one
with the number of iterations increased.

The behavior demonstrated by the $[672,336,16]$ code is different.
Looking, first, at the pseudo-codeword LP spectra we find that
configuration with the lowest effective distance is actually a
codeword, $d_{\mbox{\scriptsize ML}}=16$. We also find in the
spectrum two other codewords correspondent to $d=24$ and $d=25$.
Even thought the special low distance configurations were observed,
their frequencies were orders of magnitude smaller then of other
pseudo-codeword configurations found  at $d\gtrsim 27.33$. Emergence
of the gap suggests that, even thought the relatively small Hamming
distance will certainly dominate the largest SNR asymptotic of FER,
the moderate SNR asymptotic should actually be controlled by
continuous part of the pseudo-codeword spectra above the gap. This
prediction is indeed consistent with MC results shown in
Fig.~\ref{fig:p7}A where the early set intermediate asymptotic,
$\sim\exp(-27.33 s^2/2)$, changes to a shallower curve with the SNR
increase. Like in the case of the $[155,64,20]$ code, the
pseudo-codeword spectra of BP are qualitatively similar to the
respective one of LP.

\section{Conclusions}
\label{sec:concl}

We conclude with some general remarks highlighting directions for
future research.

One important result of this work is that analyzing LP and BP
algorithms simultaneously is helpful. The two algorithms are
asymptotically equivalent at large SNR. Currently, BP is thought of
as an algorithm of a greater practical value, however as this work
suggests, LP is easier for analysis. Therefore, developing a more
flexible, coding specific, and hopefully distributed, LP decoding of
LDPC codes may be one fruitful research direction, and some first
steps towards the challenging goals were already made
\cite{06TS,06VK,06DW,07CSa}. On the BP side of the problem, one
would,  first of all, like to develop more efficient ways of the
error-floor analysis. One conjecture here is that BP, understood as
a fixed point of BP equations, may actually be suitable for the
accelerated PCS-style analysis of the instanton spectra. In this
regards, relaxing iterative BP, e.g. through the method proposed in
\cite{06SCa}, may constitute possible resolution to the problem
caused by possibly irregular, cyclic dynamics of the iterative BP.

It was shown recently \cite{06AMU,06Amr} that codes within an
expurgated properly designed ensembles show very good convergence
(practically identical FER vs SNR dependence) in the water-fall
domain. This is in a contrary to a widely spread distribution (over
codes in the ensemble) in the error-floor domain. This important
observation of \cite{06AMU,06Amr} emphasizes importance of an
individual code analysis in the error-floor domain discussed in this
presentation. We anticipate that an ensemble approach, e.g. of the
type discussed in \cite{06AMU,06Amr}, and an individual code
instanton approach,  of the type discussed in this work, employed
together, as complementary parts of one package, could actually be
very useful in designing new efficient and application specific
coding schemes.

Let us conclude by noticing,  that instanton analysis of  the BP/LP
decoding can be easily extended to variety of interesting correlated
channels, e.g. inter-symbol-interference channel. The approach can
also be tuned to other problems in communications, storage,
operational research and network science, wherever it is necessary
to analyze algorithms of statistical inference in an extreme,  low
probability domain unaccessible to standard Monte-Carlo methods.

This work was carried out under the auspices of the National Nuclear
Security Administration of the U.S. Department of Energy at Los
Alamos National Laboratory under Contract No. DE-AC52-06NA25396.

\end{document}